\documentclass[preprint, lefttitle, number, sort&compress]{elsarticle}
\makeatletter
\def\ps@pprintTitle{%
 \let\@oddhead\@empty
 \let\@evenhead\@empty
 \def\@oddfoot{}%
 \let\@evenfoot\@oddfoot}
\makeatother
\usepackage[left=2cm,right=2cm,top=2cm,bottom=3cm]{geometry}
\usepackage{graphicx,import} 
\usepackage{natbib}
\usepackage[dvipsnames]{xcolor}
\usepackage{psfrag}
\usepackage{comment}
\usepackage{array}
\usepackage{multirow}
\usepackage{setspace}
\usepackage{amsfonts}
\usepackage{amsmath}
\usepackage{amssymb}
\usepackage{booktabs}
\usepackage{cases}
\usepackage{verbatim}
\usepackage[small,compact]{titlesec}
\usepackage{url}
\usepackage[version=4]{mhchem}
\usepackage{textcomp}
\usepackage{flafter}
\usepackage{cleveref}
\usepackage{gensymb}
\usepackage[pagewise]{lineno}
\usepackage{array}
 \usepackage{multicol}
 \usepackage{multirow}
 \usepackage{makecell}
 \usepackage{adjustbox}
\usepackage{xurl}
 \usepackage{colortbl}
 \usepackage[dvipsnames]{xcolor}
 \usepackage{subfigure}
 \usepackage{subfloat}
 \usepackage{float}
\usepackage[font=footnotesize,labelformat=simple]{subcaption}
\usepackage[toc, title]{appendix}
\newcommand{\co}{CO$_2$}
\newcommand{\PM}{PM$_{2.5}$}
\newcommand{\oC}{$^{\circ}$\textrm{C}}
\newcommand{\upth}{$^{\textrm{th}}$ }
\newcommand{\ud}{\mathrm{d}}

\newcommand\rev[1]{{\color{black}#1}} 

\definecolor{mygray}{gray}{0.6}

\begin{document} 
\begin{frontmatter}
\title{\rev{Coupled indoor air quality and dynamic thermal modelling to assess the potential impacts of standalone HEPA filter units in classrooms}
}
\author[1]{Henry C.\ Burridge\corref{cor1}}
\ead{h.burridge@imperial.ac.uk}
\author[1]{Sen Liu}
\author[2]{Sara Mohamed}
\author[1]{Samuel G.\ A.\ Wood}
\author[3]{Cath J.\ Noakes}
\cortext[cor1]{Corresponding author}
\address[1]{Department of Civil and Environmental Engineering, Skempton Building, South Kensington Campus, Imperial College London, London SW7 2BX, UK.}
\address[2]{Department of Architecture, University of Strathclyde, Richmond St, Glasgow G1 1XQ, UK}
\address[3]{School of Civil Engineering, University of Leeds, Woodhouse Lane, Leeds LS2 9JT, UK}

\begin{abstract}
The quality of the classroom environment, including ventilation, air quality and thermal conditions, has an important impact on children's health and academic achievements. The use of portable HEPA filter air cleaners is widely suggested as a strategy to mitigate exposure to particulate matter and airborne viruses. However, there is a need to quantify the relative benefits of such devices including the impacts on energy use. We present a simple coupled dynamic thermal and air quality model and apply it to naturally ventilated classrooms, representative of modern and Victorian era construction. We consider the addition of HEPA filters with, and without, reduced opening of windows, and explore concentrations of carbon dioxide  (\co), \PM, airborne viral RNA, classroom temperature and energy use. Results indicate the addition of HEPA filters was predicted to reduce \PM~ by 40--60\% and viral RNA by 30--50\% depending on the classroom design and window opening behaviour. The energy cost of running HEPA filters is likely to be only 1\%--2\% of the classroom heating costs. In scenarios when HEPA filters were on and window opening was reduced (to account for the additional clean air delivery rate of the filters), the heating cost was predicted to be reduced by as much as -13\%, and these maximum reductions grew to -46\% in wintertime simulations. In these scenarios the HEPA filters result in a notable reduction in \PM~and viral RNA, but the \co\ concentration is significantly higher. The model provides a mechanism for exploring the relative impact of ventilation and air cleaning strategies on both exposures and energy costs, enabling an understanding of where trade-offs lie.

\end{abstract}


\end{frontmatter}

\section{Introduction} \label{sec:intro}

School classrooms are ubiquitous and society's young people spend long durations within them, with the primary aim of their education and betterment; yet there is growing evidence that poor indoor conditions can have a negative impact on students' health, comfort, and academic achievements \cite{RN75,RN72}. Schools serve as a second home to students. Nearly ten million pupils in the UK attend schools, dedicating almost 30\% of their lives to this setting. While at school, they spend about 70\% of their time inside classrooms \cite{RN74,RN75,RN72}. Compared to other work and public environments, classrooms are significantly more crowded, with a density that is about four times higher than that found in office settings \cite{RN73}. 
Numerous studies have revealed that school environments often fall short in terms of quality and are generally in worse condition than office spaces and homes \cite{RN78,RN79,RN80}. This includes particular concerns around air quality, ventilation and thermal comfort. \\

Vulnerable  groups, including young children, are particularly susceptible to air pollutants \cite{RN56,RN104,RN58} and those with existing health issues, such as lower respiratory diseases—which are notably prevalent among children of primary school age—are also at heightened risk from air pollution exposure \cite{RN59}. Due to their developing immune systems and lower body weight, children are especially sensitive to environmental pollutants, as they inhale more air in proportion to their size compared to adults \cite{RN60}. Previous research indicates a correlation between exposure to particulate matter  (PM) and negative effects on the cardiovascular and respiratory systems in children \cite{RN87,RN86}. Exposure to air pollutants from indoor and outdoor sources may negatively influence primary school children's intellectual and cognitive development, potentially leading to a decline and impaired learning abilities. For example \cite{RN61} found that children attending schools in areas with high pollution levels exhibited less cognitive development growth compared to those in areas with lower air pollution levels. A meta-analysis \cite{RN81} on the impact of classroom temperature on children's school performance also shows a clear impact on students' academic achievements. Analysis of London schools revealed that air quality inside classrooms is often poorer than outdoor conditions, with concentrations of PM$_{10}$ and \PM\ surpassing WHO recommended limits \cite{RN67}. The significance of this lies in the considerable amount of time children spend in classrooms, where the levels of particulate matter PM$_{2.5}$ and  PM$_{10}$ can vary greatly due to factors like  indoor sources, and particle resuspension \cite{RN68}, as well as entry of pollutants from outdoor environments \cite{RN70}. \\ 

Classroom environments have come under greater scrutiny during the COVID-19 pandemic, with growing evidence that airborne exposure is an important transmission mechanism for the SARS-CoV-2 virus and other respiratory viruses \cite{Wang2021}. Pre-pandemic evidence suggests classroom ventilation is correlated to illness absence \cite{Mendell2013}, and several studies during the pandemic point to school ventilation as a factor in exposure \cite{Gettings2020, Buonanno2022}. Studies also suggest  the spread of respiratory infections may correlate with the levels of \PM\ particles present in classroom environments \cite{RN62,RN63,RN64,RN66}. National guidance requires adequate ventilation in schools \cite{BB101}, however several studies measuring the concentration of carbon dioxide (\co\ \!\!) as a proxy for ventilation, suggest that standards may not always be met \cite{Burridge23, Vouriot21}. A large proportion of UK schools are naturally ventilated, yet these usually rely on opening windows which can pose a substantial challenge to thermal comfort, particularly in cold or wet weather. Concerns have also been raised over the impact of opening windows on energy consumption, especially as energy prices have increased over the past two years.  

Improving ventilation is necessary in many schools, but can be costly and take time to retrofit. Air cleaning approaches are potentially a rapid intervention that can reduce exposure to both air pollutants and respiratory viruses, and can mitigate some of the challenges with opening windows for ventilation. Air cleaners employ various operating principles and filtration techniques. However, those with the highest efficiency, capable of capturing sub-micron particles, feature high-efficiency particulate air (HEPA) filters. Previous studies have demonstrated that mechanical filters, such as HEPA filtration purifiers, effectively reduce indoor pollution/particles and provide cardiovascular and respiratory advantages to individuals \cite{RN85,RN82,RN84}. Air cleaners have been installed in schools as an intervention to protect children from air pollution including in China \cite{RN94,RN89}, Korea \cite{RN90} and the UK \cite{RN92,RN94,RN91}. A study in China \cite{RN106} found that implementing the addition of air cleaners in classrooms could lead to reductions in excess of 20\% in personal \PM\ exposure. In the USA \citet{RN107} showed that using the air cleaner at its highest flow rate might raise the particle decay rate from 3.9--4.8\,h$^{-1}$ (without the air cleaner) to 6.5--6.7\,h$^{-1}$. Air cleaners also received considerable attention during the COVID-19 outbreak \cite{RN103}.\\

The performance of air cleaners is influenced by factors such as the size of the room, the device clean air delivery rate (CADR), and the placement of the air purifier within the building \cite{RN98,RN96}. There is good evidence that they mix the air within rooms and reduce the concentration of particles, including viruses. However, the effectiveness of using an air cleaner compared to the alternative approach of using natural ventilation to reduce exposure to particles and viruses has not been evaluated. In particular, the significance of combining air quality and thermal performance modelling in naturally ventilated school classrooms to evaluate the effectiveness of standalone HEPA unit air cleaners, their energy cost and their health impacts has not been extensively studied. The majority of research in schools has concentrated solely on the advantages of utilising air cleaners. Only a handful of studies have taken into account the operational costs associated with air cleaners when evaluating their impact on reducing exposure, as well as the balance between benefits and these costs.
The aim of this paper is to develop a modelling approach that:
\begin{enumerate}
    \item Proposes a simplified model for evaluating air quality and thermal comfort in classrooms, based on the assumption of ``well-mixed'' air conditions in a standard UK classroom size.
    \item Analyses the effectiveness of \rev{lowering occupants' exposures when using air cleaners, in differing intervention scenarios, and included within two different typologies of UK classrooms,} factoring in both the advantages and energy consumption control.
\end{enumerate}
The modelling approach developed aims to enable exploration of trade-offs between energy, comfort and air quality, and lays a solid foundation for optimising engineering control strategies for a healthier and sustainable classroom indoor environment.

\section{Methods} \label{sec:meth}

\subsection{Classroom HEPA model} \label{sec:model}

\rev{A fast-running coupled indoor air quality and dynamic thermal model of a classroom with the addition of standalone HEPA filters, the CHEPA model, underpins the insights presented herein.} A method to simultaneously model the air quality and thermal performance of school classrooms, naturally ventilated by both wind and buoyancy (temperature differences), is presented. The model is intended to be the simplest form capable of assessing classroom performance both in the presence, and absence, of standalone HEPA filter units. We assume that the air within the classroom is `well-mixed' such that the temperatures and pollutant concentrations are uniform, but are enabled to be different to those outdoors. The internal geometry of the classroom is approximated by a cuboid of volume $V_a = L_c \times W_c \times H_c$, where $L_c=10$\,m and $W_c=5.5$\,m are set to be the length and width of the classroom such that $L_c \times W_c=55\,\textrm{m}^2$ as is regarded as typical in the UK, e.g.\cite{BB103,Vouriot23}, and its height, $H_c$ (which is varied between classroom type, see \S\ref{sec:classr}).

We assume all airflows are incompressible and Boussinesq; such that air can be considered to be of constant density, herein $\rho_a=1.27\,\textrm{kg}\,\textrm{m}^{-3}$; provided that differences in the density of air (here arising due to temperature differences) are accounted for via incorporation of a buoyancy term within the equations of motion. Buoyancy (or reduced gravity) being defined as $\beta \, g \, \Delta T$, with $\beta$ being the thermal expansion coefficient of air (herein taken to be $\beta=1/300$\,K$^{-1}$), $g$ the gravitational acceleration and $\Delta T = T_a - T_{out}$ is the temperature difference between the classroom air and that outside. One consequence of these assumptions is that consideration of the fluxes of air masses are therefore appropriately reflected by conservation of their volume fluxes, with the equal (and opposite) inwards and outwards ventilation volume fluxes denoted $Q$.

The model classroom is assumed to have two external walls, of area $A_e = (L_c+W_c)H_c$ and effective thermal transmittance (U-value) of $U_e$. It is connected to the outdoors via windows of height $H_w$ which when fully opened have a total opening area $A_w$, with $A_w = 1.6\,\textrm{m}^2$ taken throughout. We assume that airflows, including leakage, to other indoor spaces within the school are negligible; however, we account for \rev{thermal energy} exchanges between the classroom air and the building fabric of the classroom and school buildings via incorporation of a thermal mass term with our \rev{thermal energy} equations. The effects of thermal mass are modelled via consideration of a mass $M_m = \rho_m A_m P_m$ with $\rho_m=1\,750\,\textrm{kg}\,\textrm{m}^{-3}$ the characteristic density of the thermal mass and $P_m$ the characteristic penetration thickness (of heat) within it; with exposed area of the thermal mass $A_m = (L_c+W_c)H_c + L_c \times W_c$ taken to be the area of the two internal walls and the area of the floor. Neglecting the thermodynamic effect of humidity transport, \rev{the thermal energy (here equivalent to enthalpy)} of the air within the classroom evolves, in time $t$, according to
\begin{equation}
    \rho_a \, V_a \, c_{pa} \frac{\ud T_a}{\ud t} = S_H - \rho_a \, c_{pa} \, |Q| \Delta T - U_e  A_e \Delta T - U_m A_m (T_a - T_m) \, , \label{eq:enth}
\end{equation}
with $T_a$ and $c_{pa}=1.005\,\textrm{kJ}\,\textrm{kg}^{-1}\,\textrm{K}^{-1}$ being the temperature and specific heat capacity of the classroom air, respectively, and $S_H$ the sum of all heat sources within the classroom at any instant. The temperature of the thermal mass $T_m$, with its specific heat capacity taken to be $c_{pm}=0.8\,\textrm{kJ}\,\textrm{kg}^{-1}\,\textrm{K}^{-1}$, is determined by solution of
\begin{equation}
    \rho_{m} P_m c_{pm} \frac{\ud T_m}{\ud t} =  U_m (T_a - T_m) \, . \label{eq:therm}
\end{equation}
The coupled ordinary differential equations \eqref{eq:enth} and \eqref{eq:therm} are solved via forth-order Runge–Kutta methods. In the case of \eqref{eq:enth} and \eqref{eq:therm}, the temperature of the thermal mass is initially set equal to the classroom air temperature, i.e. $T_m(0) = T_a(0)$, with the classroom air temperature initially set to the average of the outside air temperature and lower limit of the classroom temperature band, i.e. $T_a(0) = 0.5(T_{out}(0)+T_l)$.

The ventilation volume flux, $Q$, at any instant is based on the open area of the windows, the pressure coefficients and the reference wind speed, $u_{ref}$, and the buoyancy and window height. Windows may be opened some fraction $O_f$, and it is assumed that exchange flows have a loss coefficient, $C_d=0.24$ \citep[herein taken to be appropriate][]{Dalziel91}, such that the ventilating airflows experience an effective opening area $A^{*} = C_d \, O_f \, A_w$. Ultimately, the ventilation flow is determined by
\begin{equation}
    Q = A^{*} \sqrt{\beta \, g \, \Delta T \, H_w + 0.5 \, dC_p \, u_{ref}^2 } \, \label{eq:Q}
\end{equation}
where $dC_p$ is the effective wind pressure coefficient, or in the cross-ventilated case the difference in pressure coefficients between windward and leeward facing windows. We take $dC_p=0.2$ which lies well within the values expected based on the full-scale measurements of \citet{Gough19}. 

To assess the air quality, and the potential for resulting exposures, we model concentrations of exhaled \co\ concentration, $C$, exhaled viral RNA copies, $R$, and the concentration of particulate matter \PM, $P$. These are prescribed to evolve according to
\begin{equation}
     \frac{\ud C}{\ud t} =  \frac{S_C}{V_a} - \frac{Q}{V_a}(C-C_{out}) \, , \label{eq:C}
\end{equation}
\begin{equation}
     \frac{\ud P}{\ud t} =  \frac{S_P}{V_a} - \left( \frac{Q}{V_a} + \kappa_P + \frac{\eta_P \, Q_F}{V_a} \right)P + \frac{Q}{V_a}P_{out}\, , \label{eq:P}
\end{equation}
and
\begin{equation}
     \frac{\ud R}{\ud t} =  \frac{S_R}{V_a} - \left(\frac{Q}{V_a} + \kappa_R + \lambda_R + \frac{\eta_R \, Q_F}{V_a} \right) R  \, , \label{eq:R}
\end{equation}
where $S_C$, $S_P$ and $S_R$ and the sources of \co\, \PM\ and viral RNA within the classroom, $C_{out}=400\,$ppm (herein taken to be constant) and $P_{out}$ the concentrations of \co\ and \PM~within the air coming in from outside, \rev{the settling of particulate matter \PM\ and viral RNA is modelled, as is standard, via the settling rates $\kappa_P=0.4\,\textrm{hr}^{-1}$ and $\kappa_R=0.24\,\textrm{hr}^{-1}$, with the values selected based on the data reported by \citep{Kim20} and \citep{Buonanno20}, respectively; we further take} $\lambda_R=0.63\,\textrm{hr}^{-1}$\citep[see][]{Buonanno20} \rev{for} the viral inactivation/decay rate, $\eta_P=0.9997$ and $\eta_R=0.999$ \citep[see, for example,][]{Philips3000} the removal efficiencies of the HEPA filter units appropriate for \PM\ and RNA respectively, and $Q_F$ the total clean air delivery rate (CADR) from the HEPA filter units. The initial conditions, required to obtain solutions for \eqref{eq:C}, \eqref{eq:P}, and \eqref{eq:R} are $C(0)=C_{out}$, $P(0)=P_{out}(0)$, and $R(0)=0$ for the carbon dioxide, \PM, and viral RNA copies, respectively.

All parameters, that have not been prescribed constant values above, vary in time or with simulated scenario; the latter being document in the following section (\S\ref{sec:params}). With such parameterisation, in combination with the initial conditions prescribed, \eqref{eq:Q} can be solved on each time step and thereafter the system of equations \eqref{eq:enth}, \eqref{eq:therm}, \eqref{eq:C}, \eqref{eq:P}, and \eqref{eq:R}.

\subsubsection{Heating and ventilation control algorithms} \label{sec:HV}

Solutions will only yield results representative of a UK classroom if the sources of heating and the ventilation provision are appropriately parameterised and controlled. Key to achieving this is the prescription of a band of temperatures within which occupants tolerate the temperature without taking action. For simplicity, we set the low and high end of this `classroom temperature band' to be $T_l=19$\,\oC\ and $T_h=24$\,\oC\, respectively.

The school's heating system provides a heating input $E_H$ to the classroom and is set to provide heating to the classroom between the hours of 04:00 and 17:00, and only  during school days; outside of these days and times the heating system is inactive and $E_H=0$. To reflect the nature of central heating systems in many UK classroom (i.e. largely comprised of water filled radiators), changes in the system output are forced to take time to alter and $E_H$ can only change at a rate of $\ud E_H \ud t = \pm$3\,W/s up to the systems maximum heat output within the classroom, $E_{HMax}$ (see table \ref{tab:therm} for the values selected). For example, when starting up, a classroom with a heating system capable of delivering an output of $E_{HMax}=5$\,kW would take approximately 28\,minutes to reach full output with the reverse being true for its cooling down. Based on the current classroom temperature, relative to the classroom temperature band, the heating system setting is either switched, if $T<T_l$, so that the heat output is increasing or, if $T>T_h$, so that the heat output is decreasing --- each at a rate of $\ud E_H \ud t = \pm$3\,W/s. The system is then left increasing its heat output (up to $E_H=E_{HMax}$) or decreasing its heat output (down to $E_H=0$), unless $T_L \leq T \leq T_H$ is satisfied at which point $E_H$ is then held at its current value, until further switching occurs.

\vspace{\baselineskip}
The ventilation control is provided by varying the window opening extent. To reflect the manual nature of this, the window opening fraction $O_f$ is prescribed to take one of five discrete (ordered) values $O_{fs}$; the first $O_{fs}(1)$ (and lowest) set to represent the natural leakage of the classroom building envelope, with the last being unity, i.e. $O_{fs}(5)=1$, to represent fully opened windows. As the UK government has made efforts to provide \co\ monitors to each classroom in England \citep{DfEMon21}, the \co\ concentration within the classroom is also taken to provide some influence over the decisions reflected within our window opening logic --- this is enabled by first setting a `classroom \co\ band' with the concentration $C_l$ defining the low end of band and $C_h$ the high end.

At each instant, if $T<T_l$ and $C<C_h$, the window opening fraction $O_f$ is decreased to the next lower value of $O_{fs}$ (unless $O_{f}=O_{fs}(1)$ already). Alternatively $O_f$ is increased to the next higher value of $O_{fs}$ (unless $O_{f}=O_{fs}(5)$ already) if either $T>T_h$ or $C<C_l$. To reflect the fact that occupants are unlikely to be constantly checking the windows, $O_f$ remains constant if the window opening fraction has been changed within the last 15\,minutes.

\subsection{Classroom and scenario parameterisation} \label{sec:params}

\subsubsection{Occupancy} \label{sec:occparams}

The presence of occupants within classrooms provides the very motivation for our study; not only do we investigate occupant exposures, occupants provide significant sources of heat, \co\, \PM\, and potentially virus too. We prescribe the impact of occupants to be the same on each school day irrespective of classroom type or scenario. We define the number of occupants at any instant by $N$. On each school day, we vary the occupancy linearly with two people (representative of staff) arriving between 07:00 and 08:00 (such that $N=2$ at 08:00), 20 children then arrive between 08:00 and 09:00, with the classroom fully occupied thereafter --- with $N=N_{max}=32$ --- until the lunch hour when the occupancy decreases to $N=10$ at 12:00; the arrival of occupants at the beginning of the school day is reflected in their leaving at the end of the day with all pupils having left by 16:00 and then two staff leaving between 16:00 and 18:00. Outside of these hours and on non-school days $N=0$. 

On average, each occupant is prescribed to: produce a heat input of $E_O = 60$\,W \citep[e.g.][]{Vouriot23}; breathe at a rate $p=8.82\times10^{-5}\,\textrm{m}^{3}\textrm{s}^{-1}$; produce \co\ at a generation rate $G_C=3.35\times 10^{-6}\,\textrm{m}^{3}\textrm{s}^{-1}$\citep[see][for details]{Persily17,Vouriot23}; and generate \PM\ (via breathing and other activities) at a rate $G_P=0.139\,\mu\textrm{g}\,\textrm{s}^{-1}$\citep{Kim20}. As such, the sources of \co\ and \PM within the classrooms are determined by $S_C = N \, G_C$ and $S_P = N \, G_P$, respectively.

To predict the effect of the HEPA filter units on transmission of respiratory infections we consider the SARS-CoV-2 virus that causes COVID-19. Following \citet{Vouriot21} and \citet{Burridge22}, we model the number of infected occupants $I$ such that when the room is fully occupied a single individual is present, i.e. $I=N/N_{max}$. We report the number of viral RNA copies, RNA$_{C}$, via their average per-person production based on the concentration within an infected person breath, $C_R = 2.65\times10^5\,\textrm{RNA}_C\,\textrm{m}^{-3}$ \citep[see][]{Morawska09,Buonanno20}; giving the RNA source term, $S_R = I \, p \, C_R$.

\subsubsection{Incidental heat loads}

 Incidental heat loads associated with electric lighting and other electrical equipment are included via the term $E_E$. Following the classroom ICT equipment heat gains calculator \citep[see the BB 101 calculation tool,][]{BB101ToolICT}, we take the heat load associated with pupil and staff ICT devices, data projector, etc., to be 270\,W and we take that of all other electrical devices (e.g. lighting, etc.) to average out at 130\,W. Therefore during occupied hours we take $E_E = 400$\,W and outside of occupied hours we take the residual load to be $E_E=40$\,W.

\subsubsection{HEPA filter units and classroom \co\ bands}

When HEPA filter units are present, they are operational only on school days between the hours of 07:30 and 16:30 during which times they provide a CADR of $Q_F = 0.25\,\textrm{m}^3\,\textrm{s}^{-1}$ (equivalent to 900\,m$^{3}$\,hr$^{-1}$) and consume $E_F = 60\,$W of electricity (which is ultimately dissipated as heat into the classroom); otherwise, $Q_F=0$ and $E_F=0$. The parameters assumed for the HEPA filter units correspond to the specifications of the two largest trials/roll-outs of HEPA filters within UK classrooms; namely the Class-ACT study \citep{Noakes23} and HEPA filter initiative currently being led by Hertfordshire County Council \citep{HertsHEPA}.

The guidance provided during the pandemic, e.g. \citet{SAGE21}, suggests that \co\ concentrations below 800\,ppm might be indicative of good ventilation and those above 1\,500\,ppm might ``indicate overcrowding or poor ventilation and mitigating actions are likely to be required'' --- we therefore set two threshold values, $C_1=800$\,ppm and $C_2=1\,500$\,ppm, which are used to inform the classroom \co\ band. The scenario of there being no HEPA filter units present constitutes our first scenario set and our `Baseline cases'; here the classroom carbon dioxide band is defined by $C_l=C_1$ and $C_h=C_2$. In the second scenario set, termed `HEPA', we prioritise the reduction of the spread of airborne infections and assume that any existing ventilation provisions and behaviours remain unchanged, and that the CADR of the HEPA filter units be regarded as entirely additional; here again the classroom carbon dioxide band is defined be $C_l=C_1$ and $C_h=C_2$. 

\rev{In the final scenario set, termed `HEPA Adjusted' or `HEPA Adj', we imagine that one wishes adjust ventilation behaviours due to the presence of the HEPA filter unit and support some reduced natural ventilation provision by the CADR of the HEPA filter units. This might be inspired by the intention of both reducing respiratory infection but also reducing the heating energy demand due to uncontrolled heat loss through natural ventilation. Within the `HEPA Adj' scenario, the classroom \co\ band is set by $C_l - C_{out} = (C_1-C_{out}) / \textrm{max}\{0, 1 - Q_F(C_1-C_{out})/(N_{max} G_C) \}$ and $C_h - C_{out} = (C_2-C_{out}) / \textrm{max}\{0, 1 - Q_F(C_2-C_{out})/(N_{max} G_C) \}$, respectively; such that the threshold values are deemed to be representative of the steady state condition and then adjusted by the full CADR of the HEPA filter units (note that the denominator being conditioned to take only positive values simply avoids contradictory values for $C_l$ and $C_h$ in cases when the CADR is greater than the steady ventilation rate required to achieve the threshold values under a steady state condition). In the `HEPA Adj' scenarios investigated herein, the adjusted \co\ thresholds are always well in excess of 5\,000\,ppm and, as such, are effectively inconsequential.}

\subsection{Simulations and scenario parameterisation} \label{sec:sim}

\subsubsection{Modern construction and Victorian era classrooms} \label{sec:classr}

\rev{The UK has a range of classrooms each with different architecture, including ventilation openings, and thermal properties. It is not possible, nor desirable to reflect the full spectrum of UK classrooms within this analysis. Instead, we consider two large classroom types: one with a design and thermal properties of a classroom of relatively `modern construction', the other broadly representative of a classroom originating from (or following) a design of the Victorian era. Whilst the Victorian era refers to much of the 19\upth century, until 1901, and one could regard `modern' classroom construction as referring to those built in the last couple of decades we use these terms broadly to reflect our analysis of classroom constructions that represent classrooms of two very different extremes present in the UK. Classrooms within the two classes differ greatly: Victorian classrooms have large floor to ceiling heights, relatively large glazed areas and windows, poor thermal insulation, relatively large thermal mass, and require relatively large heating loads; by contrast, modern construction classrooms are the opposite, as highlighted in table \ref{tab:therm}}.

With the floor area fixed, as that specified in \citet{BB103}, the classroom volume $V_a$, and the areas of its external wall areas $A_e$ and thermal mass $A_m$ are then set by the classroom height, see \S\ref{sec:model}. With this, and the predetermined values specified in \S\ref{sec:model}, the thermal properties of the classroom and the air within are completed upon specification of the characteristic penetration thickness, (of heat) $P_m$, within the thermal mass, and the U-values of the external wall and the thermal mass. The U-values of the external walls are taken to depend on the thermal properties of the wall fabric, determined by $U_f$, and that of any glazing, $U_g$, and its area $A_g$ according to $U_e = (A_e-A_g)U_f + A_g\,U_g$. Values to determine all remaining thermal properties of the classrooms are presented in table \ref{tab:therm} with the glazed taken area set to be 30\% of the external wall area, i.e. $A_g=0.3A_e$ \citep[see, for example,][]{Cundall14}, and the thermal properties of glazing taken to be representative of modern double glazing in the modern construction classroom and single-pane windows in the Victorian era classroom.

\begin{table}[]
    \centering
    \begin{tabular}{c c c c}
       Parameter/Classroom & Modern construction & Victorian era & Units\\
       \hline
       $H_w$ & 1.0 & 1.3 & [m]\\ 
       $H_c$ & 2.7 & 4.0 & [m]\\ 
       $A_e$  & 41.85 & 62.00 & [$\textrm{m}^{2}$]\\
       $U_f$ & 0.3 & 1.6 & [$\textrm{kW}\,\textrm{m}^{-2}\,\textrm{K}^{-1}$] \\
       $A_g$  & 12.55 & 18.60 & [$\textrm{m}^{2}$]\\
       $U_g$  & 2.0 & 5.0 & [$\textrm{kW}\,\textrm{m}^{-2}\,\textrm{K}^{-1}$]\\
       $U_e$  & 0.81 & 2.62 & [$\textrm{kW}\,\textrm{m}^{-2}\,\textrm{K}^{-1}$]\\
       $P_m$  & 0.10 & 0.33 & [m]\\
       $U_m$  & 7.70 & 2.33 & [$\textrm{kW}\,\textrm{m}^{-2}\,\textrm{K}^{-1}$]\\
       $E_{HMax}$  & 5.0 & 8.0 & [kW]
    \end{tabular}
    \caption{Values of the various parameters taken to prescribe the thermal and ventilation performance of the two classroom types examined.}
    \label{tab:therm}
\end{table}

\subsubsection{Outdoor conditions and the resulting solar gains} \label{sec:out}

Thus far, we have described a set of six scenarios per simulation (three scenarios based on the presence/usage of the HEPA filter units: \{No HEPA/baseline cases, HEPA, HEPA Adj\}; each simulated for two classroom types: \{Modern construction, Victorian era\}). To examine the potential impacts of the inclusion, and operation, of HEPA filter units within UK classroom we first validate the model using a full year weather data \citep[][Test Reference Year, TRY, data providing timeseries for $T_{out}$, $u_{ref}$, and the solar irradiance]{TRYLon,TRYLeeds}, and measurements of ambient particulate matter, $P_{out}$, for both London and Leeds \citep[sourced from][taking the data for 2023 from the urban background station in Bexley, London and Central Leeds]{AURN}, see \S\ref{sec:res}. We examine the results of these year-long simulation and then compare them to results using consistent `Specified' outdoor conditions broadly representative of a week during school term during both winter and then summer; therein the outdoor temperatures are modelled to vary sinusoidally between minimum temperatures, $T_{OutMin}$, at 00:00 and maximum temperatures, $T_{OutMin}$, at 12:00. Further details of these Specified conditions are provided below, and in table \ref{tab:scen}; for the particulate matter \PM\ outdoor concentrations within the Specified scenarios were taken to be constant at $P_{out} = 6\,\mu g \, \textrm{m}^{-3}$.

The TRY data includes appropriate values for the global irradiance, due to the direct and diffuse sunlight, that impacts a horizontal plane, $G_{I}$, in each of the two cities for each hour. These global irradiance values were then modified by a factor, calculated to be $\eta_h=0.42$, to adjust the gain from a horizontal plane to the outside of a (vertical) window (using the average value of those for North, South, East, and West facing windows) based on the data of the \citet{TM37} Technical Memorandum on solar shading, which further provided a value for the transmittance of double glazing (often referred to as a G-value) of $\eta_g=0.56$. For simplicity, these factors were not varied between scenarios nor classrooms such that the solar gains within were determined by $E_S = \eta_h \eta_g G_{I}$. Values for $G_{I}$ within the Specified scenarios were taken to be zero outside daylight hours with the start of the daylight hours set by the time $D_S$ and ending at the time $D_E$; within these hours $G_{I}$ increased to a maximum a prescribed maximum $G_{IMax}$ before then decreases back down to zero --- the variation within daylight hours was determine by one half-cycle of a sinusoid.

Outdoor conditions for \co\ and RNA were taken to be constant at $C_{out}=400$\,ppm and zero, respectively, for all scenarios. 

\begin{table}[]
    \centering
    \begin{tabular}{c | c c | c c}
       Parameter/Scenario & London & Leeds & Summer & Winter \\
       \hline
       Duration\;\;[Days]  & 365 & 365  & 7 & 7 \\
       Time step\;\;[s]  & 180 & 180 & 60 & 60 \\
       
       Outdoor temp data  & TRY data & TRY data & Sinusoidal & Sinusoidal \\
       $T_{OutMin}$\;\;[$^{\circ}$C]  & N/A & N/A & 10 & 2 \\
       $T_{OutMax}$\;\;[$^{\circ}$C]  & N/A & N/A & 20 & 12 \\
       
       Solar irradiance data  & TRY data & TRY data & Semi-sinusoid & Semi-sinusoid \\
       $G_{IMax}\;\;[\textrm{W\,m}^{-2}$]  & N/A & N/A & 900 & 400 \\
       $D_S$\;\;[time]  & N/A & N/A & 05:00 & 08:00 \\
       $D_E$\;\;[time]  & N/A & N/A & 21:00 & 16:00 \\
       
    \end{tabular}
    \caption{Values relevant to the simulations and the parameterisation of outdoor temperatures and solar irradiance for the various scenarios reported on herein.}
    \label{tab:scen}
\end{table}

\subsection{The \rev{thermal energy} gain term}

Given the above, we can now write the source term in the \rev{thermal energy} equation \eqref{eq:enth} within the classroom as the sum of the heat gains from the heating system, occupants, incidental gains (from equipment \& lighting), the HEPA filter units, and the solar gains $E_S$. That is, the source for the \rev{thermal energy} gain is given by
\begin{equation}
    S_H = E_H + N \, E_O + E_E + F \, E_F + E_S \, ,
\end{equation}
\rev{where, as above, $E_H$ is the school's heating system provides a heating input into the classroom, and $N \, E_O$, $E_E$, $F \, E_F$, and $E_S$ are the heat gains associated with any occupants, electrical equipment, HEPA filter units and solar gains, respectively.}

\section{Validation} \label{sec:val}

As discussed (\S\ref{sec:sim}), we initially test that the model and its parameterisation provide reasonable results by simulating a full calendar year in the presence of outdoor conditions generated by Test Reference Year weather files and measurements of particulate matter \PM\ from official monitoring stations.

\begin{figure}
    \centering
    \includegraphics[width=0.7\textwidth]{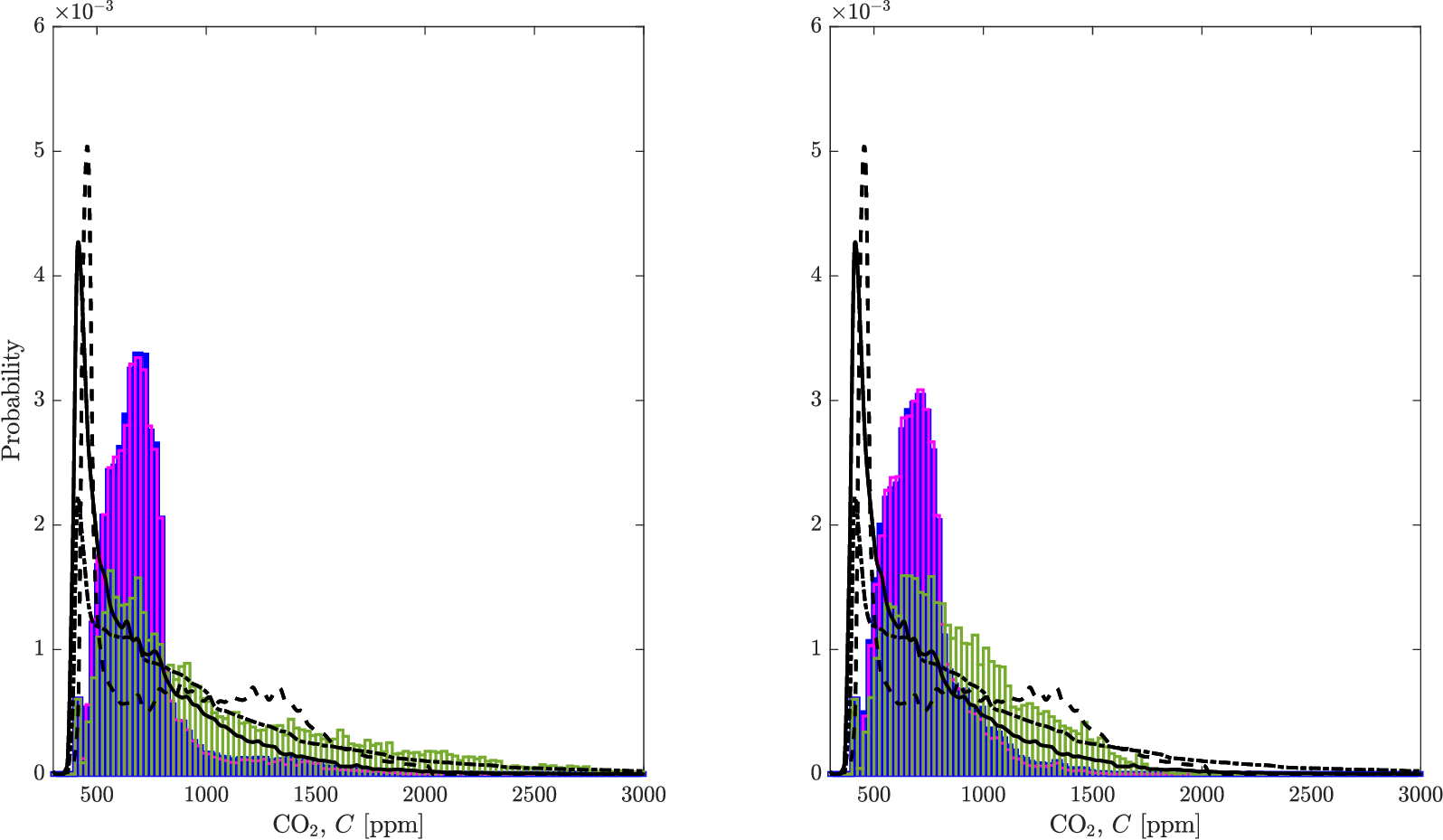}
    \caption{Histograms of the \co\ concentrations within the simulated classroom using the \rev{input} data \rev{appropriate for the city of} Leeds, UK: the left-hand pane shows the histograms of the modelled data from the Modern construction classroom and the right-hand pane those from the Victorian era classroom; both panes include black curves illustrating the PDFs from the measured data (the solid curve marks data reported within \citet{Burridge23}, the dashed curve that reported by \citet{Vouriot21}, and the dashed-dotted line marks data from the SAMHE Project \citep{Chatzidiakou23}). \textcolor{blue}{blue} histograms illustrate the baseline cases, the \textcolor{magenta}{magenta} histograms mark the HEPA cases, and \textcolor{OliveGreen}{green} the HEPA Adjusted cases.} 
    \label{fig:pdfC}
\end{figure}

\begin{figure}
    \centering
    \includegraphics[width=0.7\textwidth]{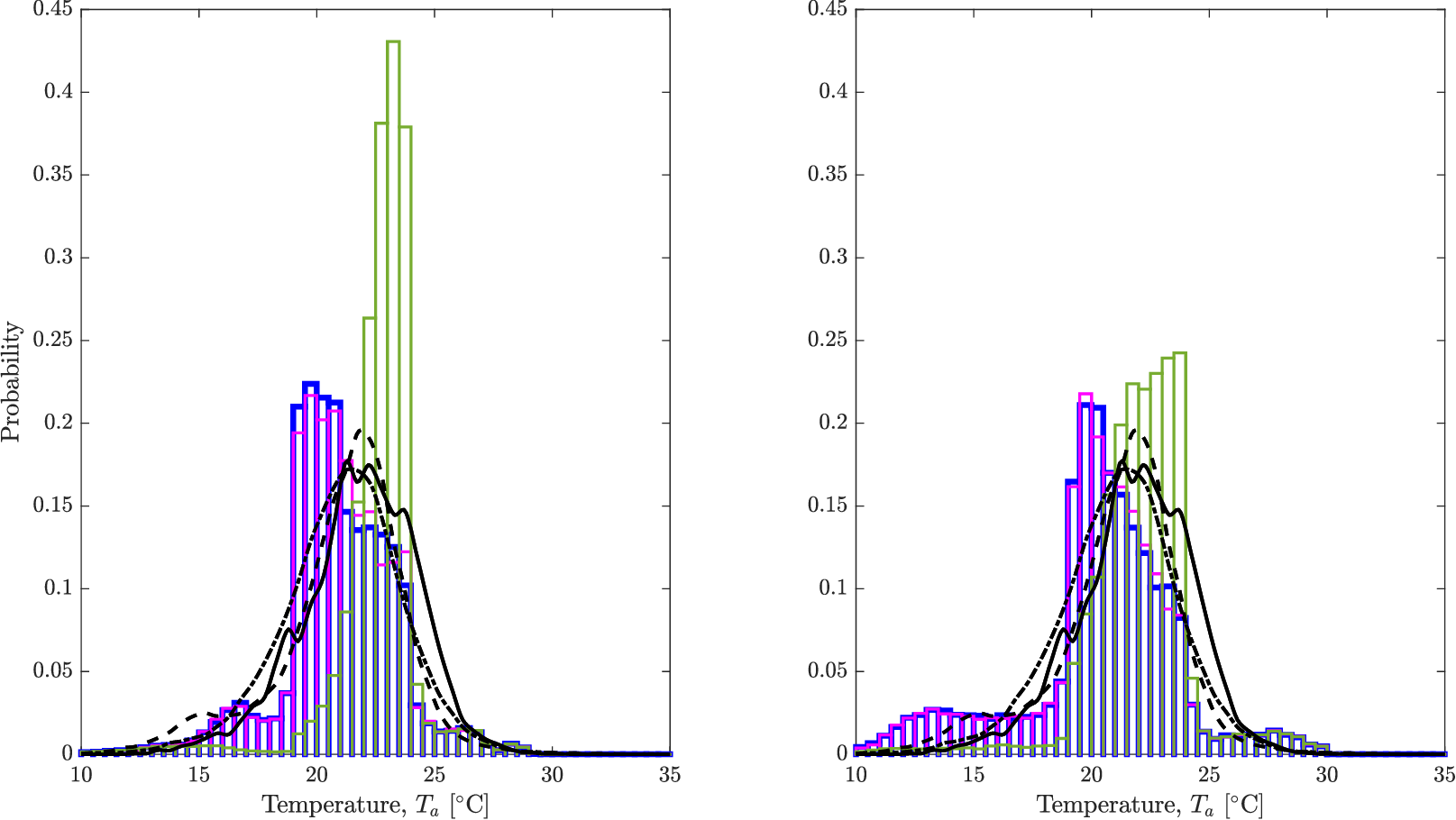}
    \caption{Histograms of the temperatures within the simulated classroom using the \rev{input} data \rev{appropriate for the city of} Leeds, UK: the left-hand pane shows the histograms of the modelled data from the Modern construction classroom and the right-hand pane those from the Victorian era classroom; both panes include black curves illustrating the PDFs from the measured data (the solid curve marks data reported within \citet{Burridge23}, the dashed curve that reported by \citet{Vouriot21}, and the dashed-dotted line marks data from the SAMHE Project \citep{Chatzidiakou23}). \textcolor{blue}{blue} histograms illustrate the baseline cases, the \textcolor{magenta}{magenta} histograms mark the HEPA cases, and \textcolor{OliveGreen}{green} the HEPA Adjusted cases.}
    \label{fig:pdfT}
\end{figure}

Using our access to data measured in UK classrooms enables comparison of the distributions of temperature and \co\ both monitored in operational classrooms and simulated with our model. The data reported by \citet{Vouriot21} provides comparison to measurements made in 45 classrooms, within 11 different schools (8 primary and 3 secondary), spanning the period November 2015 to March 2020 (i.e. pre-pandemic data); the data reported by \citet{Burridge23} enables comparison to measurements made in 36 classroom, within 4 different schools (2 primary and 2 secondary), spanning the period March 2021 to December 2021 (we take only the data from September 2021 to December 2021 as the effects of the pandemic were beginning to ease); whilst the data gathered by the SAMHE Project \citep{Chatzidiakou23} during the Autumn term of the academic year 2023-2024 provides measurements provides measurements gathered by around 300 of the SAMHE monitors deployed to schools. Comparison of the measured and simulated probability density functions (PDFs) are shown in figures \ref{fig:pdfC} and \ref{fig:pdfT} for \co\ and temperature, respectively; within, the data measured within operational classrooms are marked by black curves, and histograms present modelled data --- coloured blue for the baseline case, magenta for the HEPA case, and green for the HEPA Adjusted case (we return to discuss these latter two cases in \S\ref{sec:res} but for now note that, for both \co\ and temperature, the magenta data from the HEPA case is almost identical to the blue histogram of the baseline case). The histograms from our baseline cases (blue) show \rev{reasonable} agreement to the measured data (black curves) for temperature; the data for \co\ also shows good agreement at moderate to high concentrations but the measured data shows a sharper peak in concentration at a value about 200\,ppm lower than the baseline simulations --- to some extent, these differences may arise because the data of \citet{Vouriot21} includes classrooms with some degree of mechanically controlled ventilation. In addition \citet{Burridge23} suggested that their data was likely to have been affected by the pandemic causing more cautious ventilation behaviours within the classrooms and hence lower \co\ concentrations. \rev{Before continuing, we note that each of the six histograms representing the modelled data in figure \ref{fig:pdfC} (with the equivalent findings for the temperature data shown in figure \ref{fig:pdfT}) in each case represents results of a year-long simulation for a single classroom. By contrast, the PDFs from the three measurement campaigns, included in each of these figures, represent data gathered over multiple classrooms and over varied durations; hence, one should not expect quantitative agreement. Statistical tests of these distributions (including t-test for the means, Wilcoxon for the medians, F-test for the variances, Kolmogorov–Smirnov for distribution similarity, and Kullback–Leibler for distribution divergence) support this fact, e.g. the tests all indicate that each modelled distribution is significantly different to the distribution from each measurement campaign. Crucially, however, the tests all also indicate that the distribution of data arising from each measurement campaign is significantly different to that from the other two measurement campaigns. Differences between the measured and modelled data are not substantially larger than differences between campaigns of measured data. This suggests that, while none of these measured distributions are perfectly represented by the modelled data, the modelled data may still give a relatively good depiction of the physical condition that it models.}

Although, figures \ref{fig:pdfC} and \ref{fig:pdfT} show data from the simulations based on \rev{the input (i.e. weather and \PM) data for the city of Leeds}, the data generated using the equivalent data for London are broadly similar, e.g. see table \ref{tab:AnnEnv} for an indication. \rev{We further note that the validation has greatly benefited from the authors' access to large-scale data sets of \co and temperature within operational UK classrooms. Unfortunately, equivalent data set for \PM\ and RNA concentrations are not readily available; to compensate, we contrast our modelled \PM\ and RNA data to relevant results reported in the literature. Our modelled reductions in \PM\ concentrations (see tables \ref{tab:AnnEnv} and \ref{tab:sum}) lie well within the limits provided by the studies \citet{Duill23} and \citet{RAWAT24}, and crucially fall within the range of reductions (therein -46\% to -65\%) reported of the measurements in operational classroom of \citet{KUMAR23}. Our modelling further considers the removal of viral RNA, contained within respiratory aerosols, by HEPA filters following similar governing mechanisms to those that lead to the removal of \PM\. For viral RNA associated with SARS-CoV-2, the effectiveness of the removal mechanisms associated with HEPA filters has been established and validated by studies including \citet{Ueki22} and \citet{Parhizkar22}; as such, one can expect the validity of our modelled \PM\ reductions to hold for viral RNA.}

 We further validate our simulations via examination of the total energy within the classroom over the year. Table \ref{tab:AnnEnergy} shows the results for the simulated classrooms, gain selecting data generated using the weather and \PM\ data from Leeds. The data shows good agreement with published estimates; for example our simulated Victorian era classrooms consume around the same energy in heating as that estimated for the average school building \citep{MOHAMED21}, with our modern construction classrooms consuming around 30\% less --- perhaps it should be expected that classrooms, which are typically amongst the smaller rooms within schools with high heat gains (e.g. gains associated with classroom occupancy, etc.), exhibit lower energy consumption than the average over the whole school buildings as reported by \citet{MOHAMED21}. We therefore regard our simple classroom model as capable of generating realisable environmental conditions and energy consumption scenarios.

\begin{table}[]
    \centering
\begin{tabular}{cccc}
\textbf{Scenario --- Leeds} & Total energy [kWh/sqm] & Heating [kWh/sqm] & HEPA energy [kWh/sqm] \\ 
\hline 
Mod. Constr. --- Leeds & 260.5 & 137.9 & 0.0 \\ 
HEPA & 264.1 & 139.6 & 1.9 \\ 
HEPA Adj & 246.2 & 119.7 & 1.9 \\ 
\arrayrulecolor{mygray}\hline
Victorian era - Leeds & 344.8 & 198.6 & 0.0 \\ 
HEPA & 346.2 & 198.1 & 1.9 \\ 
HEPA Adj & 346.3 & 196.2 & 1.9 \\
\arrayrulecolor{black}\hline
 & Equip energy [kWh/sqm] & Solar Gain [kWh/sqm] & Occupant energy [kWh/sqm] \\ 
\hline 
Mod. Constr. --- Leeds & 25.5 & 49.0 & 46.2 \\ 
Victorian era --- Leeds & 25.5 & 72.6 & 46.2 \\ 
\arrayrulecolor{black}\hline
\\
\textbf{Scenario --- London} & Total energy [kWh/sqm] & Heating [kWh/sqm] & HEPA energy [kWh/sqm] \\ 
\arrayrulecolor{black}\hline
Mod. Constr. --- London & 262.3 & 137.7 & 0.0 \\ 
HEPA & 263.0 & 136.5 & 1.9 \\ 
HEPA Adj & 256.0 & 127.6 & 1.9 \\ 
\arrayrulecolor{mygray}\hline
Victorian era --- London & 350.7 & 201.7 & 0.0 \\ 
HEPA & 351.0 & 200.0 & 1.9 \\ 
HEPA Adj & 345.8 & 192.8 & 1.9 \\ 
\arrayrulecolor{black}\hline
 & Equip energy [kWh/sqm] & Solar Gain [kWh/sqm] & Occupant energy [kWh/sqm] \\ 
\hline 
Mod. Constr. --- London & 25.5 & 50.9 & 46.2 \\ 
Victorian era --- London & 25.5 & 75.4 & 46.2 \\ 
\arrayrulecolor{black}\hline
\end{tabular}
    \caption{Annual energy consumption for the simulated classrooms in Leeds and London, expressed as kilowatt-hour per square meter of floor area. Selected contributions to the energy consumption are displayed; note that the gains from (non-HEPA) equipment, solar, and occupants do not alter with the addition, nor operation, of the HEPA filter units.}
    \label{tab:AnnEnergy} 
\end{table}

\section{Results} \label{sec:res}

\subsection{Year-long duration simulations} \label{sec:year}

Irrespective of whether one examines the simulations based on the data appropriate for Leeds or London, the simulations of the Victorian era classroom suggest that around 200\,kilowatt-hours per square metre is required annually to heat the classroom in the baseline case, with the Modern Construction classrooms requiring around 30\% less energy for heating. The addition of HEPA filter units adds, just less than, 2\,kilowatt-hours per square metre in all cases. For the `HEPA' scenarios, in which the ventilation algorithm makes no account for the presence of the HEPA filter units, the heating is little changed (by no more than $\pm 1.7$\,kWh/sqm, see table \ref{tab:AnnEnergy}); as is the case for the average temperature and \co\ concentrations within classrooms (see table \ref{tab:AnnEnv}). However, for the `HEPA Adj' scenarios, in which the adjusted \co\ thresholds are effectively inconsequential, table \ref{tab:AnnEnergy} shows the energy consumed by the heating is reduced by $10-20$\,kWh/sqm for the modern construction classroom and $2-10$\,kWh/sqm for the Victorian era classroom, whilst these classrooms are all $1.5-2.0$\,\oC\ warmer --- these result from the average ventilation rate of outdoor air being around 30\% less with the average \co\ concentrations increased by around 300\,ppm. 

The air quality data for the baseline cases show that the average RNA copies are around 10\% higher, and the \PM\ concentrations are about 10\% lower, for the classrooms using Leeds weather data than those using the London data (see table \ref{tab:AnnEnv}). Differences in the outdoor \PM\ concentrations in the two cities are one factor for the \PM\ results, but so too are the higher ventilation rates in the London simulations (higher, on average, by up to 10\%) --- perhaps resulting from the slightly higher solar gains and outdoor temperatures appropriate for London. This higher flow rate would act to dilute the RNA present, but would bring more \PM\ in from outdoors. However, this complexity is difficult to disentangle from within these year-long simulations and we return to explore these aspects when examining the results of the simulations with Specified conditions, see \S\ref{sec:Specified}.

What is more important to draw from table \ref{tab:AnnEnv} is the potential impact of the HEPA filter units on the classroom air quality. In the `HEPA' scenarios --- for which the energy consumed for heating is little changed --- the particulate matter \PM\ is reduced, from the baseline cases, by -41\% to -43\%, and RNA copies by -42\% to -46\%. In the `HEPA Adj' cases --- \rev{in which the \co\ thresholds are adjusted to account for the CADR of the HEPA units and hence the ventilation supply is sometimes according reduced, the }energy consumed for heating was reduced by as much as -12\% --- the particulate matter reductions increase to be within the range -49\% to -52\% whilst the reductions in RNA copies decrease slightly, falling in the range -32\% to -35\%. These results suggest that the provision of HEPA filter units has the potential to both improve classroom air quality and reduce heating bills. However, particularly the results for the required heating, vary depending on whether a modern construction or Victorian era classroom is simulated and based on the local weather data. In the following section, we examine the robustness of our findings by carrying out Specified simulations of one week's duration and with more consistent outdoor conditions.

\begin{table}[]
    \centering
\begin{tabular}{cccccc}
\textbf{Scenario} & Temp, $\overline{T}_a$ [$^{\circ}$C] & Vent, $\overline{Q}$ [m$^3$\,s] & CO$_2$, $\overline{C}$ [ppm] & \PM, $\overline{P}$ [$\mu$g\,m$^{-3}$] & RNA, $\overline{R}$ [RNA$_{c}\,$m$^{-3}$] \\ 
\hline 
Mod. Constr. & 20.9 \{ 21.4 \} & 0.36 \{ 0.38 \} &  709 \{ 681 \} & 6.52 \{ 7.26 \} & 57.7 \{ 53.3 \} \\ 
HEPA & 20.9 \{ 21.5 \} & 0.36 \{ 0.38 \} &  705 \{ 680 \} & 3.75 \{ 4.27 \} & 31.1 \{ 30.0 \} \\ 
HEPA Adj & 22.7 \{ 22.9 \} & 0.24 \{ 0.30 \} & 1068 \{ 938 \} & 3.03 \{ 3.56 \} & 40.1 \{ 37.0 \} \\ 
\arrayrulecolor{mygray}\hline
Victorian era & 20.3 \{ 21.5 \} & 0.34 \{ 0.38 \} &  712 \{ 667 \} & 6.53 \{ 7.26 \} & 58.9 \{ 51.7 \} \\ 
HEPA & 20.4 \{ 21.5 \} & 0.34 \{ 0.38 \} &  708 \{ 667 \} & 3.74 \{ 4.29 \} & 31.7 \{ 29.9 \} \\ 
HEPA Adj & 22.1 \{ 22.7 \} & 0.25 \{ 0.30 \} &  895 \{ 799 \} & 3.19 \{ 3.73 \} & 38.3 \{ 35.3 \} \\ 
\arrayrulecolor{black}\hline
\end{tabular}
    \caption{Average environmental and air quality metrics within the classrooms from the year-long simulations using the weather and particulate matter data for Leeds, with values obtained using the London data included within the curly braces. All averages are taken during the times 09:00--16:00 \citep{BB101}, and on school days only.}
    \label{tab:AnnEnv} 
\end{table}

\subsection{Specified simulations} \label{sec:Specified}

\begin{figure}
    \centering   
    \includegraphics[width=0.8\textwidth]{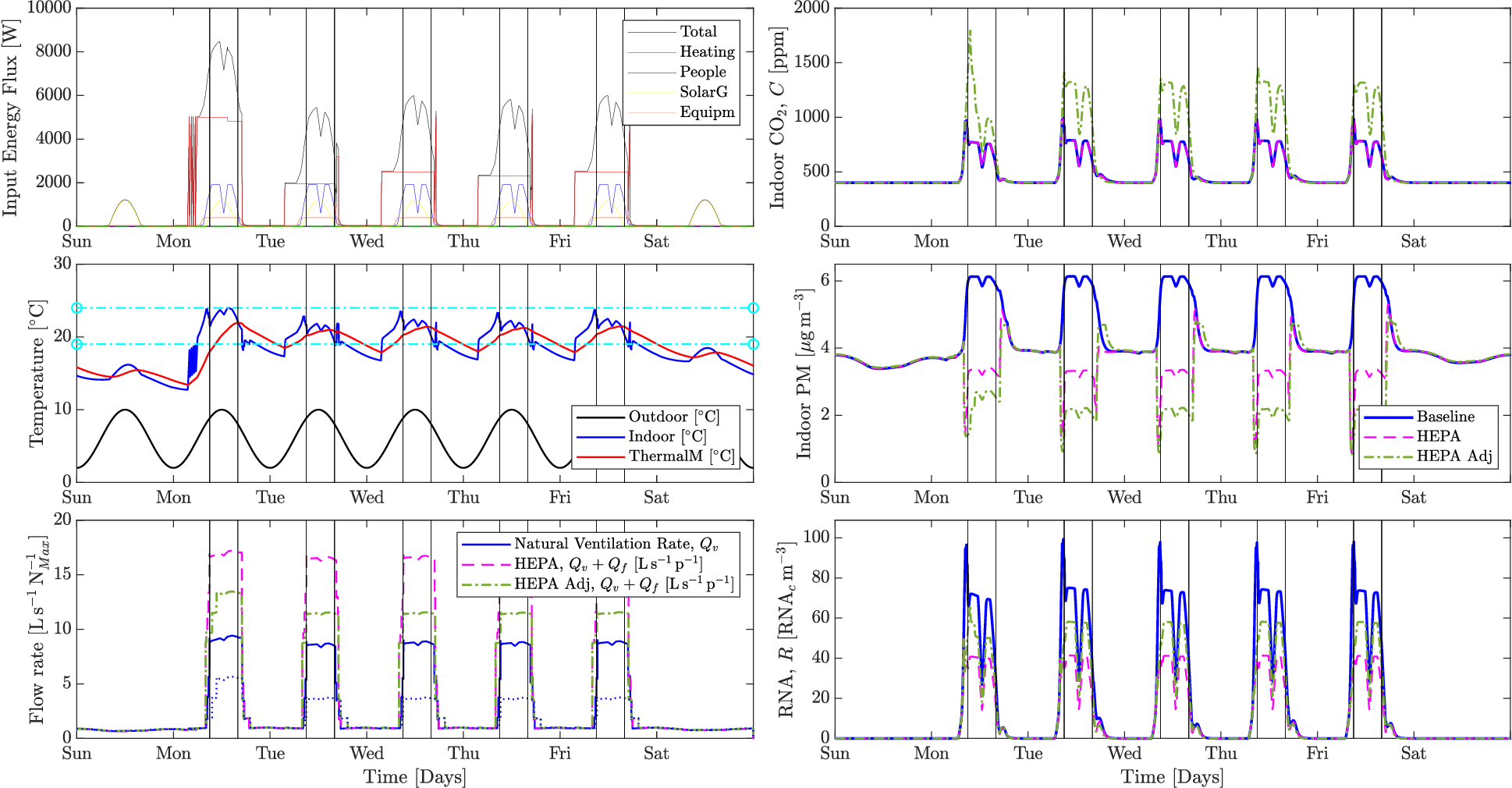}
    \caption{Simulation of winter-like conditions.}
    \label{fig:mod_wint}
    \vspace{0.5\baselineskip}
\bigskip
    \includegraphics[width=0.8\textwidth]{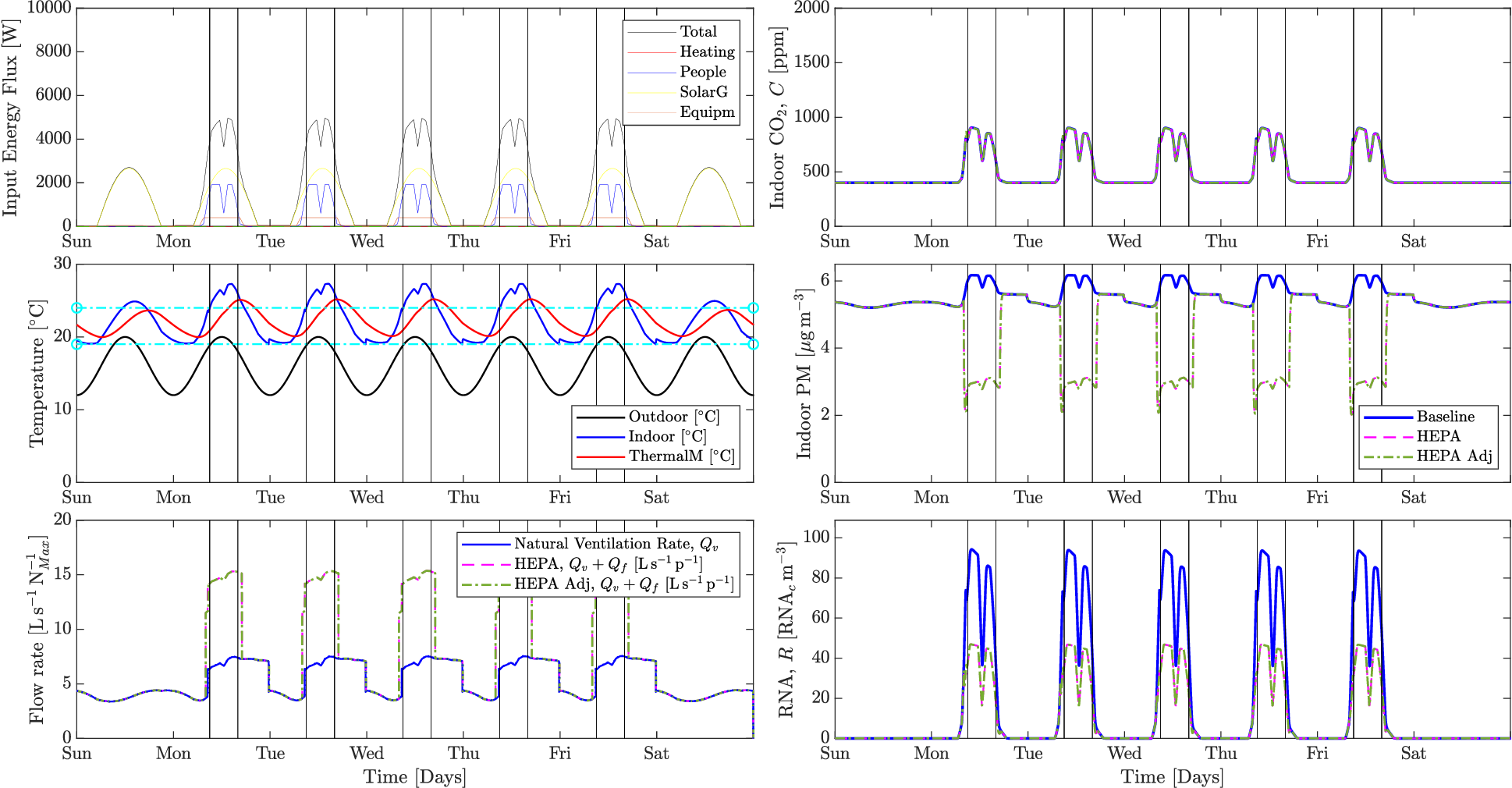}
    \caption{Simulation of summer-like conditions.}
    \label{fig:mod_sum}
    \caption{Time series data from week-long duration simulations of a modern construction classroom, during: a) winter-like conditions, and b) summer-like conditions. The top-left pane shows the energy usage within the classroom, including the energy usage broken down by a variety of sub-usages and \rev{the middle-left pane shows the temperature data --- these two panes show only data from the baseline case. The bottom-left pane shows the ventilation rate marked by blue curves, data from the baseline, `HEPA' and `HEPA Adj' scenarios are included but the ventilation rates only differ from the baseline for the `HEPA Adj' scenario in wintertime (marked by a dotted blue curve); data for total of the total of the ventilation supply and the CADR are further included. The three panes in the right-hand column also include data from the `HEPA' and `HEPA Adj' scenarios.} The top-right and middle-right panes shows the classroom \co\ and \PM\ concentrations, respectively; and the bottom-right shows the concentration of RNA copies within the classroom.}
    \label{fig:mod}
\end{figure}

The details of the conditions imposed which constitute our `Specified', one-week duration, simulations were presented in \S\ref{sec:params}. The modern construction and Victorian era classrooms, examined in \S\ref{sec:year}, remain unchanged. These shorter duration simulations allow insights from graphical exploration of the time series data with figure \ref{fig:mod_wint} showing data from a week-long simulation of winter-like conditions and figure \ref{fig:mod_sum} shows the equivalent for summer-like conditions --- we arbitrarily choose to include plots of the modern constructions classroom only, and note that the trends are broadly mimicked in the data from simulations of the Victorian era classroom. In both figures \ref{fig:mod_wint} and \ref{fig:mod_sum} the left-hand column of three panes show data only from the baseline cases, in the right-hand column data from the `HEPA' and `HEPA Adj' scenarios are overlaid within the three panes. \rev{We note again that in the `HEPA Adj' scenarios the \co\ thresholds are adjusted to account for the CADR of the HEPA units and hence the ventilation supply can sometimes be reduced, this is the case in the wintertime simulation; however, in the summertime simulation the \co\ thresholds are of no effect (due to the classroom temperatures) and the `HEPA' scenarios and the `HEPA Adj' scenarios are, in effect, equivalent as shown by the data for both of these scenarios lying coincident in figure \ref{fig:mod_sum}.}

The middle-left panes in figures \ref{fig:mod_wint} and \ref{fig:mod_sum} show the relevant temperatures within the simulations. The outdoor temperature is marked by the black curve which varies sinusoidally in both seasons, as described in \S\ref{sec:params}; cyan and magenta horizontal lines mark the lower and upper limits of the classroom temperature band, respectively. As expected, since there is no classroom cooling (air conditioning) within the model, the classroom temperature (marked by the blue curve in the middle-left panes of figures \ref{fig:mod_wint} and \ref{fig:mod_sum}) is always above the outdoor temperature, but is influenced by the outdoor temperature via heat loss across the external walls and the ventilating flow of outdoor air. This can be seen (bottom-left panes) to increase during occupied hours as windows are opened more widely in response to the classroom conditions (note that the ventilation rate attained is itself coupled to the classroom--outdoor temperature difference, with lower rates attained when the difference is smaller in summer). The classroom temperature is further influenced by the thermal mass (red curve in the middle-left panes), according to coupled equations \eqref{eq:enth} and \eqref{eq:therm}, and the heat loads --- these are illustrated in the top-left panes. Under summer-like conditions (figure \ref{fig:mod_sum}) occupancy and solar gains dominate, but with simulation of winter-like conditions the top-left pane of figure \ref{fig:mod_wint} shows that gains from the heating system, which are coupled to the classroom temperature (via the algorithm described in \S\ref{sec:HV}). All of this, in both scenarios, results in classroom temperatures that, during the school week, never drift outside the temperature bands by more than a few degrees.

The right-hand column of figures \ref{fig:mod_wint} and \ref{fig:mod_sum} depict the air quality metrics for the classrooms, with the data from the base-line case marked by blue curves. These shows elevated \co\ levels during each school day (top-right panes) that, due to a combination of the classroom's modern construction and the window openings, never exceed \co\ concentrations of 1\,000\,ppm in either scenario, and exhibit dips in concentration during the (lower occupancy) lunch hour --- these dips in concentration are evident in all three of the air quality metrics. The middle-right pane of figure \ref{fig:mod_sum} shows that outside of school hours the classroom \PM\ concentrations lie close to (but just below) the ambient level of $P_{out} = 6\,\mu g \, \textrm{m}^{-3}$ because the classroom remains relatively well ventilated even overnight; during school hours, concentration increase marginally in the presence of the occupants and their activities. Overnight \PM\ concentrations are lower under winter-like conditions as the lower nighttime ventilation rates allow the effect of settling to be more evident but during school hours, \PM concentrations attain very similar levels to those in summer. The trends in the concentration of RNA copies, noting that the prevalence of an infected person is not adjusted between scenarios, are similar with concentrations peaking between 70--90\,RNA$_{c}\,$m$^{-3}$ under both the winter- and summer-like conditions.

Most interesting to explore from the right-columns of figures \ref{fig:mod_wint} and \ref{fig:mod_sum} is the impact of the presence and operation of the HEPA filter units. Firstly, the addition of the HEPA filter units without any change in the ventilation provision of outdoor air\rev{, the `HEPA' scenario,} has no impact on the \co\ levels under either condition (see the magenta dashed curves in the top-right panes of figures \ref{fig:mod_wint} and \ref{fig:mod_sum} lie on top of the solid blue curves). This remains the case even \rev{in the `HEPA Adj' scenario in which the CADR of the HEPA filter units is accounted for within the classroom \co\ bands influence on the ventilation opening strategy (see the green curves marking the `HEPA Adj' scenario data)} since conditions are such that the windows are fully opened during the school days under summer-like temperatures. However, the same data in the top-right pane of figure \ref{fig:mod_wint} shows \co\ concentrations are further elevated in the `HEPA Adj' scenario during winter-like conditions as the CADR of the HEPA units can be used to offset a reduction in the ventilation supply of outdoor air, and hence reduce the heating needs of the classroom. Table \ref{tab:WeekEnergy} shows that for the modern construction classroom under winter-like conditions the heating needs are reduced by around -45\%, and for the Victorian era classroom the reduction is around -7\%. Moreover, during school hours --- when the HEPA filter units are operational --- the reductions in concentration of \PM\ and RNA copies are significant under both winter- and summer-like conditions. During summer-like conditions, since the presence of the HEPA filter units never alters the ventilation the results for the `HEPA' and `HEPA Adj' scenarios are identical with both \PM\ and RNA copies concentrations being reduced by about -50\% (see table \ref{tab:WeekEnv}). The results for winter-like conditions are more nuanced; the decrease in the \PM\ concentrations drops slightly to around -45\% while a decrease of around -65\% is achieved from when the CADR of the filter units is accounted for within the `HEPA Adj' scenarios. As to the concentration of RNA copies, the decreases are lower in both scenarios --- to around -45\% and -25\% for the `HEPA' and `HEPA Adj' scenarios, respectively. These results remain similar for the Victorian era classrooms. Moreover, the results for the environmental conditions within the classroom within our Specified simulations are similar to, and shed light on, the more complex trends observed within the year-long simulations using representative outdoor data. Table \ref{tab:WeekEnv} also include the exposure potentials to \PM\ and RNA copies within the classroom, these are simply the integral of the respective concentration and the breathing rate over the desired duration; here they are presented as potential exposures per day and the results align with the respective daily average concentrations.

\begin{table}[]
    \centering
\begin{tabular}{cccc}
\textbf{Scenario --- winter} & Total energy [kWh] & Heating [kWh] & HEPA energy [kWh] \\ 
\hline 
Modern Construction & 320.4 & 187.1 & 0.0 \\ 
HEPA & 324.6 & 188.6 & 2.7 \\ 
HEPA Adj & 239.4 & 100.6 & 2.7 \\  
\arrayrulecolor{mygray}\hline
Victorian era & 473.0 & 319.4 & 0.0 \\ 
HEPA & 480.3 & 324.0 & 2.7 \\ 
HEPA Adj & 455.6 & 296.5 & 2.7 \\ 
\arrayrulecolor{mygray}\hline
 & Equip energy [kWh] & Solar Gain [kWh] & Occupant energy [kWh] \\ 
\hline 
Modern Construction & 23.8 &  42.1 & 64.7 \\ 
Victorian era & 23.8 &  62.4 & 64.7 \\ 
\arrayrulecolor{black}\hline
\\
\textbf{Scenario --- summer} & Total energy [kWh] & Heating [kWh] & HEPA energy [kWh] \\ 
\arrayrulecolor{black}\hline
Modern Construction & 280.7 &   0.0 & 0.0 \\ 
HEPA & 283.4 &   0.0 & 2.7 \\ 
HEPA Adj & 286.1 &   0.0 & 2.7 \\ 
\arrayrulecolor{mygray}\hline
Victorian era & 393.1 &  21.1 & 0.0 \\ 
HEPA & 395.7 &  21.1 & 2.7 \\ 
HEPA Adj & 398.4 &  21.1 & 2.7 \\ 
\arrayrulecolor{black}\hline
 & Equip energy [kWh] & Solar Gain [kWh] & Occupant energy [kWh] \\ 
\hline 
Modern Construction & 23.8 & 189.5 & 64.7 \\ 
Victorian era & 23.8 & 280.7 & 64.7 \\ 
\arrayrulecolor{black}\hline
\end{tabular}
    \caption{Weekly energy consumption for the classrooms under the Specified simulation conditions; the top section shows data from the simulations with winter-like conditions, those from the summer-like conditions are shown below. Selected contributions to the energy consumption are displayed; note that the gains from (non-HEPA) equipment, solar, and occupants do not alter with the addition, nor operation, of the HEPA filter units.}
    \label{tab:WeekEnergy} 
\end{table}

\begin{table}[]
    \centering
\begin{tabular}{cccccc}
\textbf{Scenario} & Temp, $\overline{T}_a$ [$^{\circ}$C] & Vent, $\overline{Q}$ [m$^3$\,s] & \co, $\overline{C}$ [ppm] & \PM, $\overline{P}$ [$\mu$g\,m$^{-3}$] & RNA, $\overline{R}$ [RNA$_{c}\,$m$^{-3}$] \\ 
\hline 
Winter & 21.9 \{ 20.2 \} & 0.279 \{ 0.277 \} &  731 \{ 732 \} & 6.05 \{ 6.06 \} & 63.4  \{ 63.6 \} \\ 
HEPA & 22.0  \{ 20.3 \} & 0.280 \{ 0.280 \} &  730 \{ 731 \} & 3.28 \{ 3.27 \} & 34.9 \{ 34.9 \} \\ 
HEPA Adj & 23.1 \{ 22.1 \} & 0.124 \{ 0.184 \} & 1169 \{ 938 \} & 2.18 \{ 2.66 \} & 48.4 \{ 42.5 \} \\ 
\arrayrulecolor{mygray}\hline
Summer & 26.2 \{ 27.6 \} & 0.224 \{ 0.264 \} &  808 \{ 742 \} & 6.07 \{ 6.06 \} & 76.3 \{ 65.5 \} \\ 
HEPA & 26.3 \{ 27.7 \} & 0.225 \{ 0.265 \} &  806 \{ 741 \} & 2.98 \{ 3.21 \} & 38.6 \{ 35.7 \} \\ 
HEPA Adj & 26.3 \{ 27.7 \} & 0.225 \{ 0.265 \} &  806 \{ 741 \} & 2.98 \{ 3.21 \} & 38.6 \{ 35.7 \} \\ 
\arrayrulecolor{black}\hline
\end{tabular}
\begin{tabular}{ccc}
\\
Scenario & PM exposure [$\mu$g Day$^{-1}$] & RNA exposure [RNA$_{C}$ Day$^{-1}$] \\ 
\hline 
Winter & 13.5 \{ 13.5 \} & 141 \{ 141 \} \\ 
HEPA &  7.3 \{ 7.3 \} &  78 \{ 78 \} \\ 
HEPA Adj &  4.8 \{ 5.9 \} & 108 \{ 94 \} \\ 
\arrayrulecolor{mygray}\hline
Summer & 13.5 \{ 13.5 \} & 170 \{ 146 \} \\ 
HEPA &  6.6 \{ 7.1 \} &  86 \{ 79 \} \\ 
HEPA Adj &  6.6 \{ 7.1 \} &  86 \{ 79 \} \\ 
\hline 
\end{tabular}

    \caption{Average environmental \& air quality metrics, and exposures from simulations under the Specified conditions; data shown for the Modern Construction classroom, with the data from Victorian era classroom included within the curly braces. All averages and exposure potentials are taken during the times 09:00--16:00 \citep{BB101}, and on school days only.}
    \label{tab:WeekEnv} 
\end{table}

\subsection{Predicted exposures in classrooms and extensions to infection risk modelling}

\begin{figure}
    \centering
    \includegraphics[width=\textwidth]{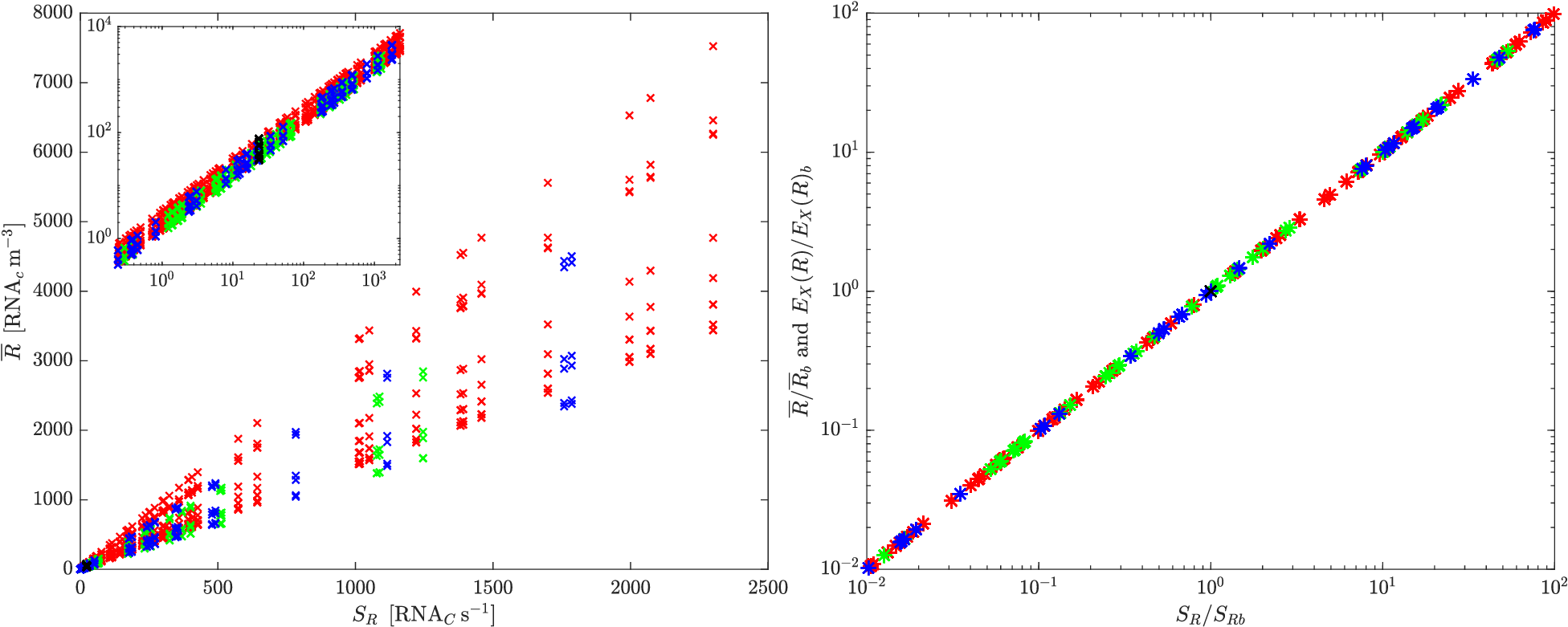}
    \caption{The variation in average concentration of RNA copies, and potential exposures, with the RNA source term, $S_R = I \, p \, C_R$. In each figure black crosses, $\times$, mark the results from the set of our 24 reported simulations and scenarios (i.e. for three HEPA based scenarios in both a Modern Construction and Victorian era classrooms, each simulated with year-long simulations using data for London and for Leeds, and Specified winter and summer conditions); all other data plotted, mark results from the 24 simulations and scenarios with differing values taken for the RNA source term. The RNA source termed was randomly varied over four decades of magnitude and the red data marks 80 different RNA source term values each ran for the Specified winter and summer conditions, green data denotes results for the 40 values ran with year-long simulations using data for London, and blue data the 30 values using data for Leeds. The left-hand pane shows the data unscaled, within the same data is inset plotted on log-log axes. The right-hand pane shows the same data and includes results for the potential exposure to RNA, denoted $E_X(R)$ (and marked by plus signs `+'), from each simulation --- all data within is scaled on the appropriate values from the corresponding reported simulation and scenario, denoted with the subscript $`b'$.} 
    \label{fig:varSr}
\end{figure}

The left-hand pane of figure \ref{fig:varSr} plots the variation in average concentration of RNA copies, with the RNA source term, $S_R = I \, p \, C_R$, within the simulated classrooms. Data from a total of 1\,404 simulations and scenarios are included, with the RNA source term taking 153 different values, pseudo randomly varied over four orders of magnitude centred on the RNA source term value (see, \S\ref{sec:params}) used within the reported simulations and scenarios. The left-hand pane plots the values of the average concentration of RNA copies, $\overline{R}$, with the inset showing the same data on log-log axes. The concentration of RNA copies varies by more than four orders of magnitude with variations in the RNA source term, and changes in the particular simulation and scenario set-up. For example, at any given value of the source term $S_R$, the average concentration of RNA copies varies by a factor of two, depending on the simulation and scenario.

The same data is plotted in the right-hand pane of figure \ref{fig:varSr}, with the total exposure potentials, $E_X(R)$ also plotted, within each simulation and scenario. However, in the right-hand pane all data has scaled on the appropriate values from the corresponding reported simulation and scenario, i.e. the scaled by the results for that simulation and scenario set-up but with $S_R$ taking the value reported within \S\ref{sec:params}. All data is seen to collapse over the range to the single line, $E_X(R)/E_X(R)_b = \overline{R}/\overline{R}_b = S_R/S_{Rb}$ with the subscript $`b'$ denoting the reported simulations and scenarios. This result is not unexpected since the coupled ODEs being solved are linear and the control of the classroom environment is independent of the concentration of RNA copies (deliberately so since we currently lack the means to measure and report such values to occupants). However, this result is significant since it illustrates that for any given simulation and scenario tested, one can \textit{a posteriori} calculate the concentrations and exposures to RNA, and more generally to any indoor source which can be expected not to influence the control of the classroom environment, that would arise should a different value for the source have been taken. Examining the source term $S_R = I \, p \, C_R$, we see it is linear in the number of infectors within the class (the occupying population), their breathing rate, and the concentration within exhaled breath. It is worthy to consider the number of infectors within the class $I$, relative to the class size $N$, as being linked to prevalence the occupying population locally, in our reported case we took the prevalence within the classroom to be $1/32\approx 0.03$.

As to infection risk modelling, figure \ref{fig:varSr} demonstrates the potential `dose' within the classroom is linear with the source term for a given parameterisation of a given disease and its source term. Moreover, when modelling airborne infection risk via Wells-Riley based approaches \citep[see,][]{Wells55,Riley78}, the form of equation \eqref{eq:R} is identical to the quanta conservation equation, e.g. see \cite{Burridge22}, so that, when the quanta generation rate, $q$, is taken to be constant (as is typical in most cases), equation \eqref{eq:R} can be multiplied by $q/C_R$ so that the source term then relates to quanta generation rate. As such, and given the understanding confirmed by consideration of figure \ref{fig:varSr}, one can re-express the results of any simulation generated by the model \textit{a posteriori}, i.e. manipulate the RNA source term to concern the quanta generation rate and/or alter its magnitude, to determine the potential `dose' arising from any airborne disease prevalence and parameterisation, or quantify all of the inputs required to directly model the airborne infection risk using Wells-Riley based approaches.

\section{Discussion} \label{sec:disc}

\begin{table}[]
    \centering
\begin{tabular}{cc|cc|c|cc}

 &  & \multicolumn{2}{c|}{Change wrt to baseline heating} & Change in & \multicolumn{2}{c}{Change in} \\
Scenario & Simulation & Heating & HEPA & Excess \co & \PM & RNA \\
\hline 
HEPA & 1-Yr (Leeds) & 1\% \{0\%\} & 1.4\% \{1.0\%\} & -1\% \{-1\%\} & -42\% \{-43\%\} & -46\% \{-46\%\}\\
HEPA & 1-Yr (Lon) & -1\% \{-1\%\} & 1.4\% \{0.9\%\} & 0\% \{0\%\} & -41\% \{-41\%\} & -44\% \{-42\%\}\\
HEPA & Winter & 1\% \{1\%\} & 1.4\% \{0.8\%\} & 0\% \{0\%\} & -46\% \{-46\%\} & -45\% \{-45\%\}\\
HEPA & Summer &  ---  &  --- & 0\% \{0\%\} & -46\% \{-47\%\} & -45\% \{-45\%\}\\
\arrayrulecolor{mygray}\hline
HEPA Adj & 1-Yr (Leeds) & -13\% \{-1\%\} & 1.4\% \{1.0\%\} & 119\% \{61\%\} & -54\% \{-51\%\} & -31\% \{-35\%\}\\
HEPA Adj & 1-Yr (Lon) & -7\% \{-4\%\} & 1.4\% \{0.9\%\} & 92\% \{49\%\} & -51\% \{-49\%\} & -31\% \{-32\%\}\\
HEPA Adj & Winter & -46\% \{-7\%\} & 1.4\% \{0.8\%\} & 133\% \{63\%\} & -64\% \{-56\%\} & -24\% \{-33\%\}\\
HEPA Adj & Summer &  ---  &  --- & 63\% \{0\%\} & -56\% \{-47\%\} & -33\% \{-45\%\}\\
\arrayrulecolor{mygray}\hline
\\
HEPA & \textbf{Average}& \textbf{0\% \{0\%\}}& \textbf{1.4\% \{0.9\%\}}& \textbf{-1\% \{0\%\}}& \textbf{-44\% \{-44\%\}}& \textbf{-45\% \{-45\%\}}\\
HEPA Adj & \textbf{Average}& \textbf{-22\% \{-4\%\}}& \textbf{1.4\% \{0.9\%\}}& \textbf{102\% \{43\%\}}& \textbf{-56\% \{-51\%\}}& \textbf{-29\% \{-36\%\}}\\
\end{tabular}

    \caption{Summary of the relative impacts across scenarios, classrooms and simulations when adding HEPA filters with no adjustments to ventilation (HEPA), and with the opening of windows being adjusted to reflect the CADR of the HEPA filter units (HEPA Adj). All values are percentage change relative to the baseline in each case for Modern Construction classrooms, with the values within curly parenthesise being those for Victorian era classrooms. Energy data for the summer simulations are omitted since these are either zero or inconsequential, see table \ref{tab:WeekEnergy} showing the energy consumed by the HEPA units is equal in summer and winter. The last two rows present simple averages across the four simulation types examined. NB: all values concerning \PM\ and RNA apply equally to daily average and to exposures.}
    \label{tab:sum} 
\end{table}

The results of the 24 simulations and scenarios are summarised in table \ref{tab:sum}, expressed as a percentage change from the results from the baseline classroom scenario, i.e. without any HEPA filter units having been provided. Results for the energy consumption of the HEPA filter units are reported with respect to the energy consumption of classroom heating --- these results are excluded for the simulations of summertime conditions since the heating consumption was either zero, or inconsequentially small.

Results for the HEPA scenario, where the opening/closing of windows was altered due to the presence of the HEPA filter units, are relatively simple to interpret: changes in the heating consumption are always small (never more than 1\%) and the cost of running the HEPA filter units is 1\%--1.5\% of the heating cost; \co\ remains largely unchanged; and both \PM\ and RNA are reduced significantly (reductions of consistently between -40\% and -50\%). However, the results for the HEPA Adjusted scenarios, in which the CADR of the HEPA units can be used to offset a reduction in the ventilation supply of outdoor air, are more nuanced. Since, on average, there is less ventilation provided to the classrooms in these scenarios the of concentrations excess of \co\ increase significantly, but the reductions in RNA concentration decrease in magnitude (typically to between -30\% and -40\%) due to the HEPA filters. Since there are both indoor and outdoor sources of \PM\ the magnitude of the reductions increase in the HEPA Adjusted scenarios (typically to between -50\% and -60\%). The impact of adding HEPA filters and considering their CADR to be a factor in determining the opening and closing of windows, i.e. the HEPA Adjusted scenarios, is the most subtle result and varies significantly between the Modern construction and Victorian era classrooms. Modern construction classrooms might be expected to reduce the energy consumption of heating by between -5\% and -15\%; Victorian era classrooms might reduce heating energy costs by around -5\% or just simply offset the running  of the HEPA filter units themselves.

The results presented here are for a simple model in a small number of scenarios\rev{; for example, the natural degradation of the filters was neglected, implicitly assuming that schools replace the filters within the HEPA units prior to this degradation becoming significant. However, the scenarios modelled} do consider key factors including classroom type, geographical location and window behaviours. However, as a modelling study there are a number of limitations which are important to note. As a zonal model, the simulations do not incorporate detail of spatial variability which would affect the actual ventilation flow patterns, pollutant concentration and thermal comfort experienced in real settings. In reality there would likely be greater variability in some of the parameters, particularly around ventilation behaviours which will depend more on individual preferences and understanding than can be incorporated in the model. Similarly the relative indoor and outdoor \PM\ sources, the particular classroom design and occupancy, and the selection of particular HEPA filter units and their usage would all affect school specific findings. Nevertheless, the results give a good insight into the relative impact of different parameters and where the trade-offs are likely to lie when selecting approaches for improving air quality in school classrooms.  

\section{Conclusions} \label{sec:conc}

The \rev{classroom HEPA model presented herein, termed CHEPA}, is a suitably fast coupled IAQ and dynamic thermal model of classrooms in the presence, and otherwise, of standalone HEPA filter units. \rev{The model is the simplest representation of the dynamics of the classroom environment required to enable estimates the potential impacts of HEPA units, see \S\ref{sec:disc} for a discussion.} The outputs from the model compare well with data measured in classrooms. The addition of HEPA filter units was predicted to reduce \PM\ concentrations by around -40\% to -60\%, depending on the classroom type and its operation in the presence of HEPA filter units --- these predictions are broadly inline with measurements made in operational classrooms within the UK \cite{Duill23,RAWAT24,KUMAR23,Noakes23} --- reductions in the concentrations of RNA copies were predicted to fall in the approximate range -30\% to -50\%. It is noteworthy that the reductions in \PM\ were higher when the ventilation to the classroom was reduced, by taking into account the CADR of the HEPA filter units within decisions regarding the opening and closing of windows; however,the opposite was true of RNA concentration with these seeing the greatest reduction when the ventilation of the classroom remained unaltered by the presence of the HEPA filter units. Moreover, the strategy of accounting for the CADR of the HEPA filter units within decisions regarding the opening and closing of windows also offered the potential for significantly reduce (1\%--13\%) the energy consumption of the classroom heating in their presence. \rev{However, accounting for the CADR of the HEPA units and (sometimes) reducing the ventilation provision accordingly, can result in elevated \co\ within the classrooms, particularly in wintertime. \co\ is widely regarded as a good indicator of classroom ventilation \citep[e.g. see][]{FINNERAN24}, and an adequate indicator of indoor air quality \citep{Lowther21}; as such, very careful consideration should be given to the scientific and practical implications if one is ever to recommend that the ventilation provision to classrooms can sometimes be reduced to account for the CADR of HEPA filters units, herein our `HEPA Adj' scenarios --- for clarity, these scenarios are not inherently recommended by the authors.}

Assuming gas heating, and taking the current energy price caps within the UK \citep[£0.29 per kWh for electricty and £0.07 per kWh for gas][]{OFGEM}, the year-long simulations predict that in the baseline case each classroom cost £550--£770 to heat (primarily dependent of classroom type). The electricity costs of running the HEPA filter units would be about £30 per year. Should the classroom ventilation be reduced, by taking into account the CADR of the HEPA filter units within decisions regarding the opening and closing of windows, then the savings on the heating costs would be predicted to fall in the range £10--£80 --- however, to achieve these savings, without exacerbating risk due to unintended consequences, would require significant education and training of classroom staff. These costs should also be viewed in the context that purchasing adequate HEPA filter units for a classroom can cost in the range of £1\,000--£2\,000\rev{, and replacement filters often cost £50--£100 with these needing replacement every year, or so}. Moreover, these units presents logistical challenges to deploy (including access to power sockets and health and safety considerations like trailing cables), take up space within classrooms, and increase noise levels within classrooms. However, consideration of their inclusion within classrooms is worthy of consideration given their potential to improve classroom air quality.

Finally, neither the CHEPA model nor this paper seeks to endorse nor promote the use of standalone HEPA filter units in classrooms --- the intention is to provide suitably accurate methods to objectively estimate the potential impacts of their deployment. If ambient air quality, particularly in urban environments, were significantly improved and schools provided with upgraded ventilation systems and strategies, then standalone HEPA filter units in classroom need not be a consideration; however, achieving either of these aspirations involves financial, time, and logistical implications that are many order of magnitudes greater than those involved in deploying HEPA filter units. As such, considered and balanced investment decisions are required by schools, their authorities, and our governments. The CHEPA model is intended to help support these decisions, which would be best achieved by deploying access to the model via a tool that is usable for schools, their authorities and governments, and provides appropriately interpretable results --- work to deliver this tool is underway.

\section*{Acknowledgements}

HCB and SGW would like to acknowledge the contribution of everyone involved in the SAMHE Project, including the schools, teachers and pupils, and thank all members of the SAMHE Steering Committee and Engagement Panel for their continued support and guidance --- they acknowledge funding from the SAMHE Project, as an extension of the CO-TRACE project, which was funded by the EPSRC under grant number EP/W001411/1, and received additional funding from the UK’s Department for Education. Further funding and support was received from the UKRI SPF Clean Air Networks, specifically: HCB, CJN and SGW received funding from the NERC Breathing City: Future Urban Ventilation Network, under grant number NE/V002082/1, and HCB received funding from the NERC Tackling Air Pollution at Schools Network, under grant number NE/V002341/1. Finally, the authors are grateful to the Chartered Institution of Building Services Engineers, in particular Technical Director Dr Hywel Davies, for permitting use of the CIBSE weather files.

\section*{Competing interests}
The authors declare no competing interests.




\bibliographystyle{elsarticle-num-names}
\bibliography{bibl}

\end{document}